\documentclass{epl}
\usepackage[utf8]{inputenc}
\usepackage[T1]{fontenc}
\usepackage[a4paper]{geometry}
\usepackage{epsfig,amsmath,amssymb,subfigure,rotating}
\usepackage{lmodern}

\newcommand{\vectr}[1]{{\mathbf{#1}}}

\title{On the connected-charges Thomson problem}
\author{An\v ze Slosar \inst{1} and Rudolf Podgornik \inst{1,2,3} }
\institute{
\inst{1} Dept. of Physics, Faculty of Mathematics and Physics, \\
University of Ljubljana,  SI-1000 Ljubljana,  Slovenia\\
\inst{2} University of California at Santa Barbara, \\
Kavli Institute of Theoretical Physics, Santa Barbara, CA 93106 USA\\
\inst{3} Dept.  of Theoretical Physics, 
J. Stefan Institute, SI-1000 Ljubljana, Slovenia}

\begin{document}

\maketitle
\vspace{-1.5cm}
\begin{abstract}
  We investigate the modifications brought about by the linear
connectivity  among charges in the classical Thomson problem. Instead of packing
with local hexagonal order intersperced with topological defects, we find charge distributions with helical
  symmetry wound around the surface of the sphere. This finding should
  have repercussions in the viral packing and macroion adsorption
  theories.
\end{abstract}
\vspace{-0.5cm}

\section{Introduction}

Electrostatics can be very unintuitive and can lead to quite
unexpected features in a variety of physical, chemical and biological
contexts \cite{Levin-1}. The problem of packing charges on a sphere,
generally referred to as the Thomson problem \cite{Levin-2},
has recently received a lot of attention and has been analysed via
detailed direct numerical evaluation \cite{Altschuler} as well as via
approximate analytical theories \cite{Nelson}. What is clear is that
for large enough number of charges the smallest electrostatic energy
configurations have packing with local hexagonal order intersperced with 
topological defects \cite{Nelson}. A possible
question here is: what happens if one linearly connects the charges and thus
considers a {\em linearly connected} Thomson problem.

The linearly connected Thomson problem should be directly relevant for packing or adsorption of 
polyelectrolytes, {\sl i.e.} charged polymers, on or to spherical surfaces.
Adsorption of polyelectrolytes to spheres has been extensively simulated in the
context of polyelectrolyte - macroion interactions in colloid physics
\cite{Simula-1}. Various configurations of adsorbed polyelectrolytes are seen,
depending on the interaction parameters and the stiffness of the chain.
Though they all correspond in one sense or another to a free energy
minimisation as in the case of histone - DNA interactions \cite{hoda}, the 
phase space is simply too big in order to attempt
some kind of systematic classification of shapes of the adsorbed
polyelectrolyte chains. An "inverted" problem to the one just
described is represented by the investigations of the nature of
packing of DNA inside small bacteriophage capsids that has recently
received a lot of attention since the nature of packing of DNA inside
the capsid determines the ejection force that can be measured directly
\cite{Simula-2}.  The only general and consistent conclusion stemming from these studies seems to be that DNA close to the capsid
wall is more ordered than deep within the capsid. Recent cryomicroscopy on epsilon15 bacteriophage \cite{king}
gives ample experimental support for this type of order gradient within the viral capsid.
At small packing fractions, where most or indeed all of DNA is expelled towards the inner surface of the
capsid, the packing problem is directly related to the linearly connected Thomson problem.

In order to shed some light onto packing of connected objects into
confined spaces with spherical topology, we performed zero-temperature
numerical investigations of the connected-charges Thomson effect.
Apart from the fundamental significance of this problem \cite{Levin-2}
it could also be seen as a model of polyelectrolyte packing either on
the outside or the inside of a sphere in the cases where the entropy
of the chain can be neglected.

\begin{figure}
\begin{center}
  \epsfig{file=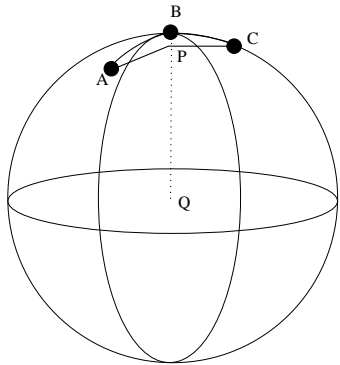,  width=0.3\linewidth, angle=0} 
\end{center}

  \caption{ This Figure illustrates the geometry used in this
    paper. Three charges labelled $A$, $B$ and $C$ lie on a surface of the sphere, whose centre
    is at $Q$. Point $P$ is such that $\overline{AP}$ and
    $\overline{PC}$ are perpendicular to the line $\overline{QB}$. The
    angles $\alpha_i$ used in this work correspond to $\pi-\langle APC$. \label{fig:slk}}
\end{figure}

\section{Model}

Our model system is composed of $N+1$ elementary charges
connected by $N$ rigid bonds of length $\ell$ on a unit 3D sphere.
Each rod subtends and angle $\phi=2\sin^{-1}(\ell/2)$ with the centre
of the sphere. The position of the charges are given by $N+1$ vectors
$\vectr{x}_i$ ($i=1...N+1$), satisfying $|\vectr{x}_i|=1$ and
$|\vectr{x}_i-\vectr{x}_{i-1}|=\ell$. These constraints reduce the
number of degrees of freedom to $N+2$. Imposing rotational invariance
further restricts the system to just $N-1$ degrees of freedom.  The
total length of the chain is $L=N\phi$. The Coulomb energy of
this system is given by
\begin{equation}
  E(L) =\sum_{i, j>i}^N \frac{1}{|\vectr{x}_i-\vectr{x}_j|}
\end{equation}
It is useful for comparison to divide this value by the minimum energy the
chain would have if it was not constrained to lie on a sphere. In that
case, the minimum energy configuration would be a completely straight
chain, whose energy is given by 
\begin{equation}
  E_{0}(L) =\sum_{i, j>i}^N \frac{1}{(j-i)\ell}=\frac{1}{\ell}\left(\Psi(N+1)+N(\gamma-1)\right),
\end{equation}
where $\Psi$ is the standard digamma function and $\gamma$ is the Euler's
constant. 

In the limit of large $N$, the minimal energy of an unconstrained Coulomb chain reduces to

\begin{equation}
    E_{0}(L) \sim 
\frac{1}{\ell} N
\log N
\end{equation}
This has a fairly obvious explanation: for a long enough chain, most
of the potential energy comes from the nearest neighbour interactions and
therefore scales as $N$, while the Coulomb interactions between more remote charges contribute
only logarithmically in $N$. This result is a signature of a completely extended Coulomb chain.

We define the reduced potential energy by 
\begin{equation}
  P(L) = \frac{E(L)}{E_{0}(L)},
\end{equation}
which measures the ratio between the Coulomb self-energy of a constrained and completely extended chain.
\begin{figure}
  \epsfig{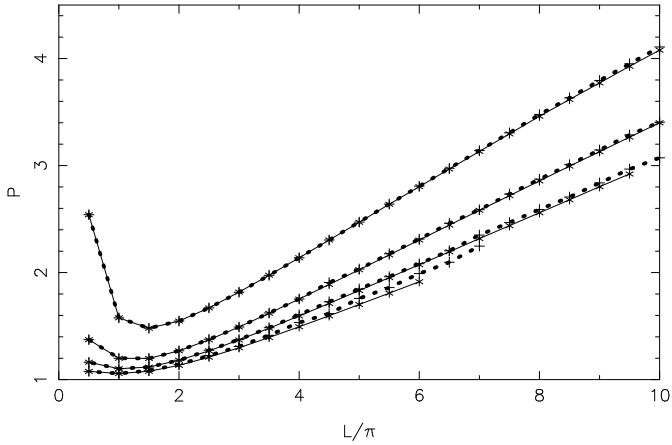}
  \caption{ This figure shows the reduced potential energy as a
    function of the total chain length $L$ for the four cases considered
    here: $\phi=\pi/4$ (top), $\pi/10$, $\pi/20$, $\pi/40$ (bottom).
    Solid line corresponds to the annealing in terms of $\alpha_i$,
    while the dotted thick line corresponds to annealing in terms of
    $\beta_i$. For large values of $N$, the computations become
    prohibitively expensive. \label{fig:12}}
\end{figure}

A natural description of this system is by $N-1$ angles $\gamma_i$
subtended between rigid rods linking $i-1$th, $i$th and $i+1$th
charge, i.e. $\cos
\gamma_i=(\vectr{x}_{i}-\vectr{x}_{i-1})\cdot(\vectr{x}_{i+1}-\vectr{x}_{i})$.
However, by projecting the mid-point down to the plane perpendicular
to $\vectr{x}_{i}$ and going through $\vectr{x}_{i-1}$ and
$\vectr{x}_{i+1}$ one arrives to a mathematically more convenient
description in terms of the angles $\alpha_i$. We will define $N-1$ angles $\alpha_i$
as

\begin{equation}
  \cos \alpha_i=(\cos \phi \vectr{x}_{i}-\vectr{x}_{i-1})\cdot(\vectr{x}_{i+1}-\cos \phi \vectr{x}_{i}).
\end{equation}
The geometry of this choice of variables is illustrated on Figure \ref{fig:slk}.

Using straightforward geometry one can show that 
\begin{equation}
  \vectr{x}_{i+1}=\cos \phi \vectr{x}_{i}+\vectr{f}_{i} \cos{\alpha_{i}} + \vectr{p}_{i} \sin{\alpha_{i}},
\end{equation}
with
\begin{equation}
  \vectr{f}_{i} =  \vectr{x}_{i} \cos{\phi}-\vectr{x}_{i-1}
\end{equation}
and
\begin{equation}
  \vectr{p}_{i} =  \vectr{x}_{i} {\times} \vectr{f}_{i}
\end{equation}
By construction $\vectr{x}_{i+1}$ lies on a sphere if both $\vectr{x}_{i-1}$
and $\vectr{x}_i$ lie on a sphere.  An $\alpha_i=0$ configuration corresponds to a linear,
completely extended chain, while $\alpha_i=\mbox{const.} \neq 0$ configuration results in charges being distributed
along an arc on a spherical surface. 

We will also use $N-1$ coefficients $\beta_i$ that can be derived simply from a
polynomial expansion of $\alpha_i$ around the midpoint of the chain
and are defined by
\begin{equation}
  \alpha_i = \sum_{j=1...N-1} \beta_j \left(\frac{i-N/2}{N/2}\right)^{j-1}.
\end{equation}
Description of the chain by $\beta_i$ or $\alpha_i$ is only a matter
of convenience.

\begin{figure}
\begin{tabular}{cc}
  \epsfig{file=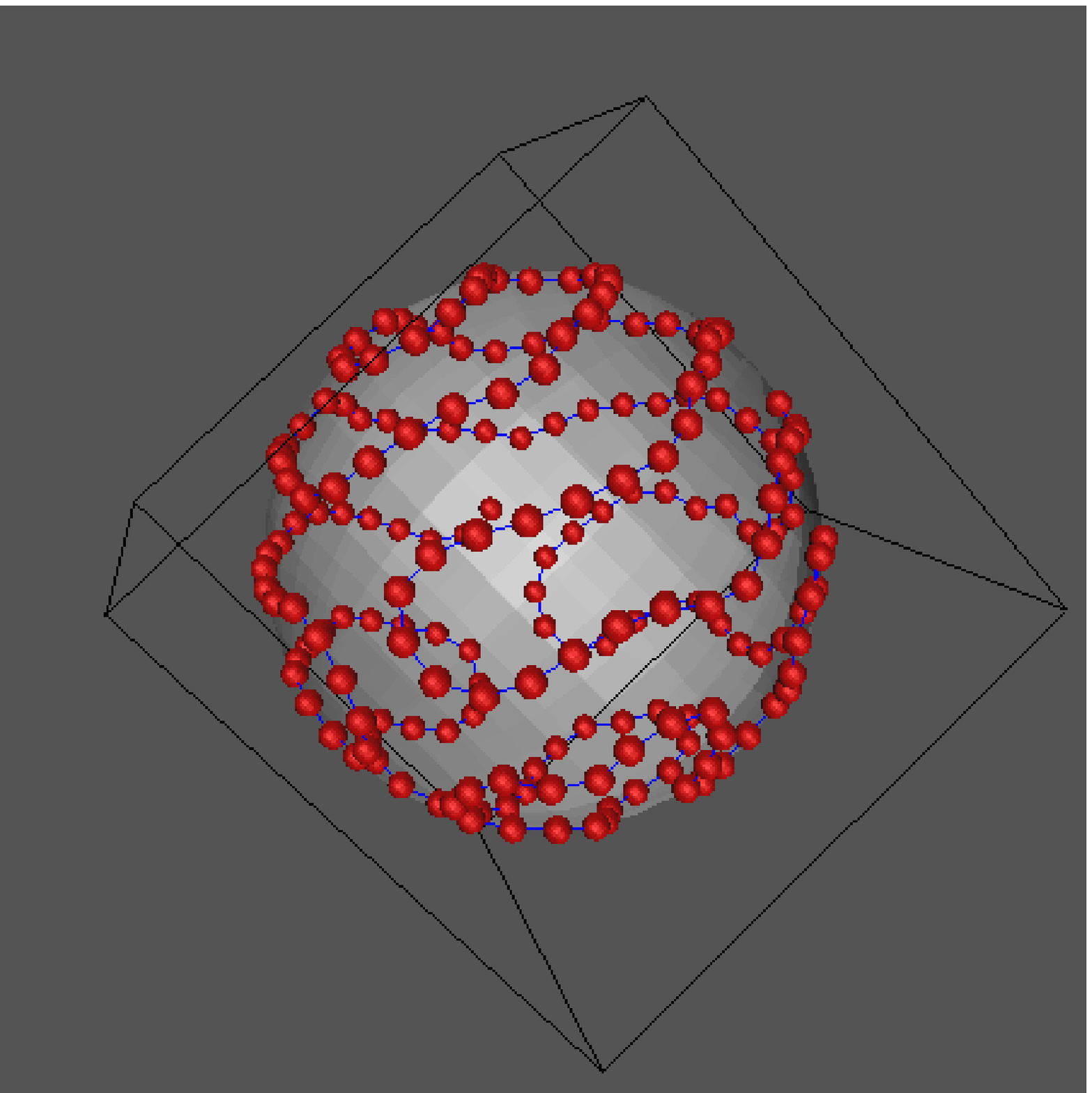,  width=0.45\linewidth, angle=0} &
  \epsfig{file=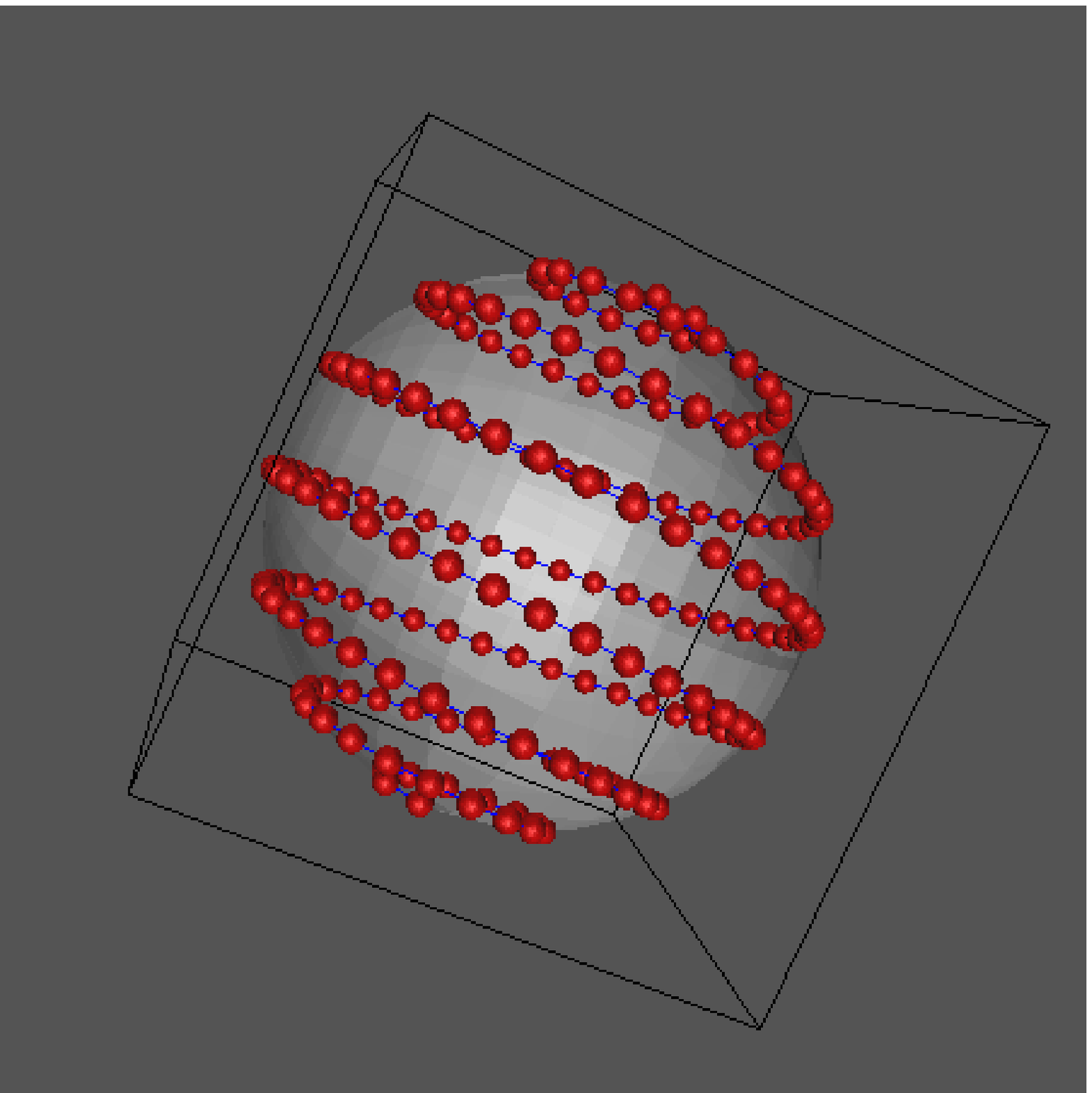,  width=0.45\linewidth, angle=0} \\
\end{tabular}
  \caption{Typical minimum energy configurations for charges on a unit
    sphere. The l.h.s. corresponds to a typical solutions obtained
    with $\alpha_i$ minimisation, while the r.h.s. is one obtained by
    $\beta_i$ minimisation. Both are very similar (within better than 0.1\%)
    in terms of the  actual energy.   \label{fig:0}}
\end{figure}

Saff and Kuijlaars \cite{Saff} have shown that the "general spiral set" can be used to generate 
uniformly distributed points on a sphere. Their formalism can not however be directly applied to 
our problem, since the fixed links constrain the system further. 
Though we are not concerned explicitly with uniform distributions 
of points on a sphere, one can, nevertheless, reasonably expect that
the minimal solution for the Thomson problem in the constrained case will exhibit 
helical behaviour. This is a hypothesis that we will try to investigate.

In order to set the terms of discussion, we recall the equation of a helix on a
cylinder as given by $z = ct$ and $\phi = t$.  This can be generalised
to a helix on a surface of a sphere by either identifying $z$ with
$\cos{\theta}$ or $\theta$. We will denote the first case as
embedded helix:
\begin{eqnarray}
 \theta &=& \cos^{-1} (ct) \nonumber\\
 \phi &=& t\label{h1}
\end{eqnarray}
and the second one as spherical helix:
\begin{eqnarray}
 \theta &=& ct \nonumber\\
 \phi &=& t\label{h2},
\end{eqnarray}
where  $\theta$ and $\phi$ are the ordinary spherical coordinates.

\section{Results}

We have programmed software that finds the minimum energy
configuration using two methods, working in either $\alpha_i$ space or
$\beta_i$ space. 

In the first case we anneal by angles $\alpha_i$ using a version of
genetic algorithm used in \cite{genetic} for a classical Thomson
problem. We start with a population of $P$ ``parents''. We create
$P(P-1)/2$ ``off-springs'' by taking first $N/2$ $\alpha_i$ values from
the first parent and the remaining $\alpha_j$s from the second parent
for all possible combinations of parents. We then use a simple
numerical minimiser (downhill simplex) to find the local potential
energy minimum corresponding to each parent. The best $P$ solutions
are kept and the process is repeated until a satisfactory convergence
is reached. We typically work with 12 initial parents and 8
generations.

In the second case we find the best solution by minimising the energy
expressed in terms of parameters $\beta_i$. Genetic algorithm is not
useful here as each $\beta_i$ affects the entire chain and so good
local features cannot be preserved. However, it is an extremely good
space for optimisation and thus finds a good solution very quickly.
After trying several options we have settled with an annealing
algorithm that is a series of linear minimisers in randomly chosen
parameters where the solution is assumed to be bracketed around the last
optimal point and the width of the bracket is decaying exponentially
until a desired convergence is reached. We typically try $\sim200$
different starting positions and then choose the best.

Finally, we have also minimised the energy of the system, whose
charges are additionally constrained to lie on either embedded or
spherical helix. Since this is a one-parameter model, the straight
numerical minimiser performs satisfactorily. The purpose of this
exercise was to check how well the true minimal configuration
can be approximated by either helix.

We have run the minimiser with four values of $\phi$: $\pi/4$,
$\pi/10$, $\pi/20$ and $\pi/40$ and for values of $L$ ranging between
$\pi/2$ and $10\pi.$

\begin{figure}
\epsfig{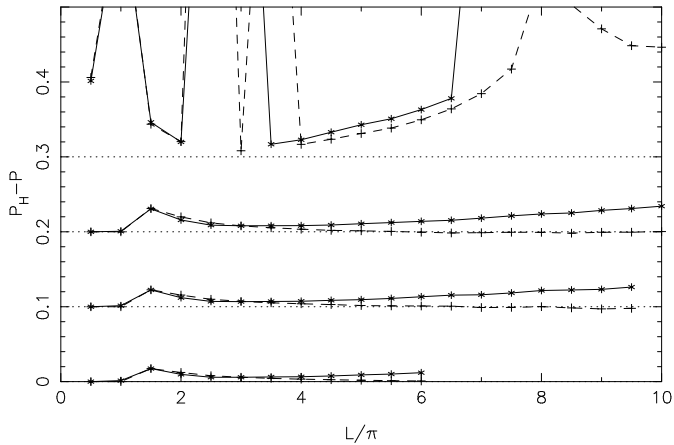} 
  \caption{This figure shows the difference in reduced potential
    energy $P$ between solutions found by full numerical minimisation
    and those found by constraining charges to lie on either embedded
    helix (solid) or spherical helix (dashed). Lines were offset
    vertically by $i\times 0.1$ and correspond the four cases considered
    here: $\phi=\pi/4$ (top), $\pi/10$, $\pi/20$, $\pi/40$ (bottom). The
    dotted lines represent the zero difference value. For large values
    of $N$, the computations become prohibitively expensive.
    \label{fig:34}}
\end{figure}

In Figure \ref{fig:12} we plot $P(L)$ as a function of the total chain
length. It is immediately clear that both annealing procedures lead to
stable solutions that are very similar in terms of energies for large enough values 
of the length of the chain. The well
discernible minimum in the $P(L)$ is easy to understand as follows: the
energy $E_0(L)$ is dominated by the nearest-neighbour contribution
to the overall potential. The penalty payed by constraining the chain
to lie on a sphere comes primarily from the ``bending of the chain''.
However, when $L \gtrsim \pi$, the $E(L)$ receives another
contribution from interaction between charges that lie far apart on
the chain but a forced to lie close in physical coordinates due to
wrapping of the chain around the sphere. Therefore, one can expect a
minimum in $P$ at around $L/\pi \sim 1$ and this is exactly what we
observe.  However, we note that for a long enough chain, the reduced
potential energy begins to scale nearly linearly with the chain
length. 

Naively, one would expect that the energy of the electrons on
the chain wrapped on a sphere would scale with  $N$ roughly as $N^2$ and so the reduced potential
energy $P$ should scale as $N/\log N$ for large $N$. Our data do not
extend high enough in $N$ to discriminate unequivocally  between linear and
$N/\log N$ scaling, though the latter indeed provides a marginally
better fit to the data. Interestingly enough, this is the true even for cases when values of
$E_0$ \emph{have not yet reached the $N\log N$ scaling}.  

We also note, that although very similar in energy, the solutions
found by the $\beta_i$ annealing procedure are always helical, while
the solutions in terms of the $\alpha_i$ annealing procedure are, for
large enough $N$, configurations that are locally (on a sphere, not
along the chain) helical but are irregular on large scales. This is
illustrated in the Fig. \ref{fig:0}. What we are likely seeing is a
plethora of metastable configurations with nearly minimal energy,
similar to those found for a classical Thomson problem. This is
supported by two observations: firstly, making minimiser convergence
criteria stronger does not improve these solutions, implying that they
are indeed meta-stable and secondly, all final generation solutions in
the genetic algorithm have very similar energies (to within a few
pro-miles) and are different on runs with different random seeds,
implying that the family of those solutions is quite large. These
nearly minimal configurations are never found by the $\beta_i$
annealing as this would require a coherent tuning of several $\beta_i$
coefficients to achieve a preferred local behaviour. The existence of
a large number of these nearly minimal configurations also implies
that unless a specific mechanism is in place, the chain is very likely
to land in a complex nearly-perfect meta-stable state rather than the
helical structure with long-scale ordering.

Next let us discuss the solutions which come from constraining the charges
to lie on either of the two helices defined in Eqs. (\ref{h2})
and (\ref{h1}). Results are plotted in the Figure \ref{fig:34}. It can
be seen that for chains with $\phi \lesssim \pi$, the spherical helix is
an increasingly better approximation to the true minimum for large enough
$L$. It thus appears that we are able to construct explicitly those configurations that are 
close to optimal.
This confirms our expectations based on the "generalised spiral set" of Saff and Kuijlaars \cite{Saff} 
that solutions with helical symmetry should describe  minimal energy states for the 
connected-charges Thomson problem.

\section{Conclusions}

In this paper we have analysed the connected-charges Thomson problem
by brute-force numerical minimisation of the potential energy. The
problem is completely specified by two parameters: $L$, the total
length of the chain, and $\phi=L/N$, that defines the inter-charge
separation. Note that the links between the charges are inextensible.

We have found that for all cases considered, the reduced potential
energy has a minimum at $L/\pi \sim 1$ and scales as $N$ or
$N\log N$ (the range of $N$s investigated does not extend to large enough $N$ so 
that one could differentiate between the two scaling forms). 
This implies furthermore that the potential energy of the chain scales as $N^2$ for large $N$ as
expected from standard arguments. However, the relation is true even for
$N$ that are so low that $E_0(N)$ has not reached the $N\log N$
scaling yet.

We have heuristically attempted to model the minimum energy
configuration with a helix on a sphere. We have found that the
cylindrical helix projected onto a sphere by a simple transformation
$z\rightarrow\theta$, thus creating a spherical helix, leads to charge configurations 
that seem to be very close to the numerically obtained minimum for $L\gtrsim \pi$ and $\phi \ll
1$.  The solution is likely to continue in such manner until the
distance between points that are far apart along the helix but close
in real space becomes important. This would result in a frustrated
helix that might hide new and exciting phenomenology. Unfortunately,
the presently available computers do not reach the computational power
necessary to answer this question, at least not with the methods
provided in this paper.

What is the physical relevance of our results? Adsorption of flexible polyelectrolytes 
onto spherical macroions (colloidal interactions) and configurations of polyelectrolytes within spherical cavities
(viral packing) both belong to the type of problems which have been analysed in this paper on their bare 
bones level. Though in general the configurations of real polyelectrolyte chains depend on many 
parameters and are difficult to classify \cite{kunze-netz}, our analysis provides an
asymptotic limiting configuration that should be discernible in the case where adsorption interaction 
onto a spherical macroion is strong enough, or the repulsive interactions along the chain in a spherical 
cavity are sufficiently long-ranged.  An especially beautiful illustration of our theory is provided by recent
elucidation of the DNA packing inside the epsilon15 capsid \cite{king} that shows tight helical winding on the
outer boundary and the progressive loosening of the orientational order as one moves away from the 
boundary towards the interior of the capsid.

\section{Acknowledgement} RP and AS acknowledge the financial support of the Agency for 
Research and Development of Slovenia under the grants P1-0055(C) and Z1-6657 respectively. 
This research was also supported in part by the National Science Foundation under Grant No. PHY99-07949.


\end{document}